\documentclass[twocolumn,prb,superscriptaddress,aps,showpacs]{revtex4}
\usepackage{graphicx}
\usepackage{dcolumn}
\usepackage{amsmath}
\newlength{\figwidth}
\newlength{\figwidthb}
\begin{document}

% \thanks{Also at Physics Department, XYZ University.}%Lines break

\title{Magnetic field dependence of charge stripe order in $\rm La_{2-x}Ba_{x}CuO_4$ ($\rm x\approx1/8$)}
\author{Jungho Kim}
\affiliation{Department of Physics, University of Toronto, Toronto,
Ontario M5S~1A7, Canada}
\author{A. Kagedan}
\affiliation{Department of Physics, University of Toronto, Toronto,
Ontario M5S~1A7, Canada}
\author{G. D. Gu}
\affiliation{Department of Condensed Matter Physics and Materials
Science, Brookhaven National Laboratory, Upton, New York,
11973-5000}
\author{C. S. Nelson}
\affiliation{National Synchrotron Light Source, Brookhaven National
Laboratory, Upton, New York, 11973-5000}
\author{Young-June Kim}
\email{yjkim@physics.utoronto.ca} \affiliation{Department of
Physics, University of Toronto, Toronto, Ontario M5S~1A7, Canada}

\date{\today}

\begin{abstract}
We have carried out a detailed investigation of the magnetic field
dependence of charge ordering in $\rm La_{2-x}Ba_{x}CuO_4$ ($\rm
x\approx1/8$) utilizing high-resolution x-ray scattering. We find
that the charge order correlation length increases as the magnetic
field greater than $\sim 5$~T is applied in the superconducting
phase (T=2~K). The observed unusual field dependence of the charge
order correlation length suggests that the static charge stripe
order competes with the superconducting ground state in this sample.
\end{abstract}


\maketitle

Charge inhomogeneity in strongly correlated electron systems such as
the cuprate superconductors has drawn much attention for its
intimate relationship with exotic charge transport properties.
\cite{Orenstein00,Dagotto05} In particular, neutron scattering
experiments have revealed that doped holes arrange themselves in
orderly rows in the copper oxide plane in certain high-temperature
superconductors. This static ordering of charge stripes and
associated spin stripes was observed in $ \rm
La_{1.6-x}Nd_{0.4}Sr_xCuO_4$ (LNSCO) crystals.
\cite{Tranquada95,Tranquada96,Tranquada99,Christensen07} More
recently, stripe ordering has been investigated in detail in $\rm
La_{1.875}(Ba,Sr)_{0.125}CuO_4$.
\cite{Fujita02b,Kimura03,Fujita04,Kimura04,Abbamonte05} In other
materials, such stripes may also exist, although in most cases the
stripes are not static and fluctuate with time and position.
\cite{Bozin00,Singer02} Understanding the role of stripe physics in
cuprate superconductors is believed to be essential in elucidating
the superconducting mechanism of the cuprates.

Despite the fundamental importance of charge ordering in the
cuprates, only a limited number of experiments have been carried out
to study charge stripes
\cite{Zimmermann98,Niemoller99,Kimura03,Abbamonte05,YJkim08} and a
comprehensive examination of the relationship between charge stripes
and superconductivity is still lacking. We note that there have been
extensive neutron scattering studies on spin stripe correlations,
including measurements of their temperature and magnetic field
dependence \cite{Katano00,Lake01,Khaykovich03,Christensen07}, while
scanning tunneling spectroscopy experiments have shown intricate
real space images of checker-board type charge distribution in a
number of cuprate superconductors. \cite{Hoffman02,Hanaguri04}
However, no systematic field dependence study of charge ordering has
been carried out to date.

We have carried out a detailed investigation of magnetic field
dependence of charge order (CO) in $\rm La_{2-x}Ba_{x}CuO_4$ ($\rm
x\approx1/8$) (LBCO) utilizing high-resolution x-ray scattering. We
find that the correlation length of the charge order exhibits
unusual magnetic field dependence. Specifically, the correlation
length increases as the magnetic field greater than $\sim 5$~T is
applied in the superconducting phase (T=2~K). The fact that the CO
correlation length grows under the field indicates that the CO
correlation length at zero field is not limited by extrinsic factors
such as disorders and defects, but is determined by microscopic
origin, such as competing ground states. We discuss the implication
of the observed field dependence of the CO correlation length to the
spatial distribution of CO and SC regions.

%Experiments

The magnetic field dependence study was carried out at the X21
beamline at the National Synchrotron Light Source (NSLS) at
Brookhaven National Laboratory. The beamline is equipped with a
superconducting magnet that can provide a magnetic field of up to
$\mu_0H$=10~T and temperatures as low as T=2~K. The incoming photon
with energy 15~keV was selected by a double bounce Si(111)
monochromator, and horizontal scattering geometry was used. In order
to reduce the background and to improve the resolution, a LiF(200)
analyzer crystal was placed in the scattering geometry after the
sample. The LBCO sample used in our measurements was grown by the
traveling-solvent floating zone technique at Brookhaven National
Laboratory. The bulk superconducting onset, $\rm T_c$, of the
particular sample studied here was measured to be approximately 6~K
from magnetic susceptibility.

%
% Figure 1
%

\begin{figure}[t]
\vspace*{-0.0cm}\centerline{\includegraphics[width=2.2in,angle=90]{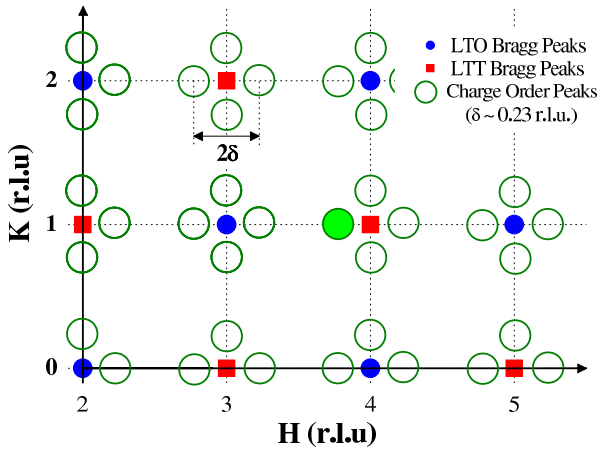}}
%\centerline{\includegraphics[width=4.5in,angle=-90]{fig1.eps}}%
\vspace*{-0.2cm}%
%\centering\epsfig{file=fig1.eps,width=5cm,angle=-90}
%
\caption{(Color online) Schematic representation of the (H,K,L)
(L=half-integer) plane of reciprocal space. Filled circles (blue)
and filled squares (red) represent LTO and LTT peak, respectively
(See the text for explanation). Large open circles (green)
correspond to the positions where CO peaks are located. Filled large
circle (green) indicates the CO peaks investigated in the current
study.}\label{fig:schematic}
\end{figure}

Fig.~\ref{fig:schematic} shows a schematic (H,K,L) (L=half-integer)
plane of the reciprocal space of LBCO. Filled circles (blue)
represent the points where Bragg peaks are found in the low
temperature orthorhombic (LTO) phase, and filled squares (red) are
the points where additional Bragg reflections occur in the low
temperature tetragonal (LTT) phase below 55~K. \cite{YJkim08} Large
open circles (green) represent the (H,K) positions of the
incommensurate superlattice peaks arising from the static CO. Note
that these CO peaks occur at the half-integer $L$ values, reflecting
the criss-crossing pattern of the stripes. \cite{Zimmermann98}
Detailed investigation of the magnetic field ($\mu_0H$) dependence
was carried out near the (3.77,1,0.5).

% Results

\begin{figure}[t]
\vspace*{0.2cm}\centerline{\includegraphics[width=3in,angle=0]{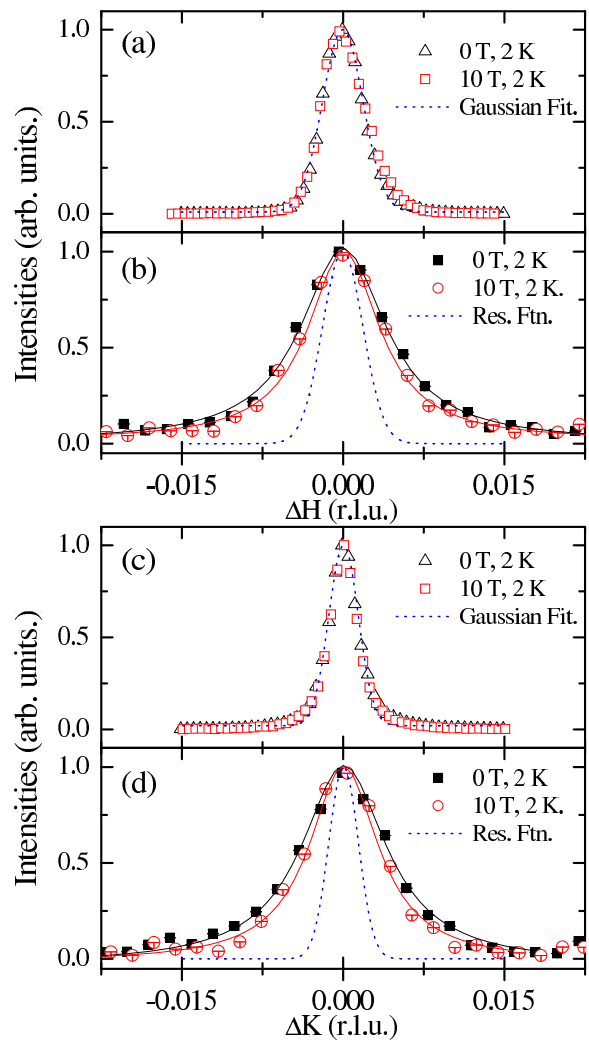}}
%\centerline{\includegraphics[width=4.5in,angle=-90]{fig1.eps}}%
\vspace*{0.0cm}%
%\centering\epsfig{file=fig1.eps,width=5cm,angle=-90}
%
\caption{(Color online) (a) and (c) show H and K scans,
respectively, of the ($\rm4~1~0$) Bragg peak for $\mu_0H$=0~T and
10~T at T=2~K. The dotted lines are Gaussian fits to the raw data.
(b) and (d) show H and K scans, respectively, of the
($\rm3.77,1,0.5$) superlattice peak for $\mu_0H$=0~T and 10~T at
T=2~K. The solid lines in (b) and (d) are fits to a Lorentizian
function convoluted with the instrumental resolution which is shown
as a dotted line. One can clearly see the sharpening of the peak in
both the H and K directions under applied magnetic field of
10~T.}\label{fig:raw}
\end{figure}

The sample was cooled in zero-field to 2~K, which is well below the
superconducting and the CO temperature of $\sim 6$ K and $\sim 42$
K, respectively. \cite{YJkim08} The magnetic field was applied
perpendicular to the $\rm CuO_2$ plane, which coincides with the
scattering plane within 2 degrees. The representative scans along
the H and K directions of the (3.77,1,0.5) peak for two magnetic
field values of $\mu_0H$=0~T and 10~T at T=2~K are shown in
Figs.~\ref{fig:raw}(b) and (d), respectively. The statistical error
bars are much smaller than the symbol size in the figure. To
facilitate easy comparison of the width change, all scans are
plotted as a function of relative momentum transfer, ($\rm\Delta H$,
$\rm\Delta K$), and are normalized to have matching peak heights.

%
% Figure 2
%

We can see that the CO peak width is not resolution limited, unlike
the Bragg peak whose width is limited by the instrumental
resolution. In fact, the width of the CO peak is about twice that of
the instrumental resolution which is shown as a dotted line in
Figs.~\ref{fig:raw}(b) and (d). In particular, one can clearly see
the sharpening of the peak in both the H and K directions under
applied magnetic field. This change is not very large (between 10
and 20 \%), but based on several tests described below, we believe
that this effect is clearly larger than experimental uncertainties.

Firstly, in Figs.~\ref{fig:raw}(a) and (c), we show similar H and K
scans for the nearby ($\rm4,1,0$) Bragg peak as a comparison. Both
the Bragg peak and the CO peak data were obtained with the same
experimental conditions in one setting. In contrast to the CO peak
widths, the Bragg peak widths in both the H and K directions are
found to remain unchanged with applied magnetic field up to 10~T.
Secondly, the measurement was repeated three times using different
combinations of magnetic fields. The first run measured the CO peak
widths at fields of 0~T and 10~T as displayed in
Figs.~\ref{fig:raw}(b) and (d), respectively. The second run
measured the peak widths at fields of 0~T, 5~T, and 10~T and the
third run at fields of 0~T, 6~T, and 8~T. After each run, the sample
was warmed above the CO temperature to eliminate the possibility of
sample history dependence. \cite{History} All three runs showed the
same magnetic field dependence of the CO peak width.

\begin{figure}[t]
\vspace*{0.2cm}\centerline{\includegraphics[width=2.5in,angle=90]{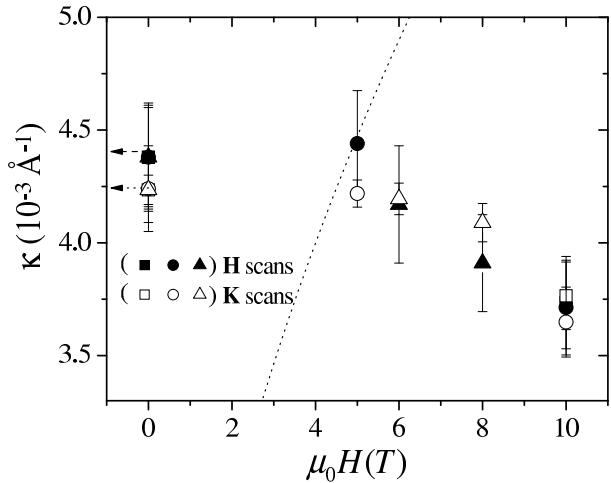}}
%\centerline{\includegraphics[width=4.5in,angle=-90]{fig1.eps}}%
\vspace*{0.0cm}%
%\centering\epsfig{file=fig1.eps,width=5cm,angle=-90}
%
\caption{The magnetic field ($\mu_0H$) dependence of the (3.77,1,0)
CO peak widths along the H and K direction. Filled and open symbols
are $\kappa_H$ and $\kappa_K$, respectively. Three different
measurements with different combinations of magnetic fields were
performed as described in the text. The dotted line is the inverse
of vortex lattice parameter ($1/a_0$) as discussed in the text.}
\label{fig:field}
\end{figure}

To see how the intrinsic peak width of the (3.77,1,0.5) superlattice
reflection changes with magnetic field, the scans shown in
Figs.~\ref{fig:raw}(b) and (d) are fitted to a two-dimensional
Lorentzian peak convoluted with the instrumental resolution which
was obtained by fitting the (4,1,0) Bragg peak to a Gaussian
function. The resulting half-width at half maximum (HWHM, $\kappa$)
values are summarized in Fig.~\ref{fig:field}. Filled and open
symbols correspond to $\kappa_H$ and $\kappa_K$, respectively. One
can immediately notice that the CO correlation is quite isotropic in
the copper oxide plane, which is due to the fact that the stripes on
neighboring layers run perpendicular to each other.
\cite{Tranquada96} The values of both $\kappa_H$ and $\kappa_K$ at
$\mu_0H$=0~T are approximately $0.0043(3)\rm\AA^{-1}$, corresponding
to a correlation length of $\sim 230\rm\AA$, which is consistent
with results reported earlier. \cite{Abbamonte05} With an increase
in $\mu_0H$, the peak width decreases, which becomes noticeable
above $\mu_0H$=5~T. At 10~T, $\kappa_H$ (or $\kappa_K$) is about
$0.0037(2){\rm\AA}^{-1}$, corresponding to a correlation length of
$\sim 270{\rm\AA}$, which is about 17\% larger than the zero-field
value. This change in the correlation lengths corresponds to the
correlation area growing from about 3500 unit cells at 0~T to nearly
5000 unit cells at 10~T, an expansion of almost 37 \%.

Our data show that the charge ordering fails to grow into a true
long range order, and its correlation length saturates at $\sim
230\rm\AA$. Extrinsic mechanisms such as compositional and
structural defects can prevent charge order from becoming a true
long range. However, the observed increase of the CO correlation
length under the field clearly indicates that the CO correlation
length is not limited by extrinsic factors, but determined by
intrinsic reasons based on microscopic physics.

%
% Figure 3
%

One should also keep in mind that the charge stripe order in this
system is accompanied by spin stripe order.\cite{Xu07} That is, in
the hole-poor region between the charge stripes, the
antiferromagnetic spin order mimics that of $\rm La_2CuO_4$. This
spin stripe order in LBCO actually goes through spin-flip transition
when about 6~T of magnetic field {\em within} the copper oxide plane
is applied.\cite{Huecker05} On the other hand, applying magnetic
field along the c-direction does not seem to affect the spin stripe
order in any significant way.\cite{lcosdw} Of course, magnetic field
along the c-direction has direct implication with regard to the
superconductivity, and indeed it was observed that there is a weak
diamagnetism below 40~K at zero field.\cite{QLi07} This temperature
at which diamagnetic signal sets in decreases with increasing field,
and the weak diamagnetism is completely suppressed around 7~T. This
suggests that the observed field dependence of CO could be the field
dependence of superconductivity, since the charge order is unlikely
to be affected by magnetic field directly.

In this regard, it is interesting to note that the ``critical" field
of $\sim 5$~T corresponds to the field at which the vortex lattice
spacing becomes comparable to the CO correlation length. Using the
available neutron scattering data for $\rm La_{2-x}Sr_{x}CuO_4$,
\cite{Keimer94,Lake02,Gilardi02} one can write the vortex lattice
spacing as $a_0 \sim 500/\sqrt{\mu_0H}$ in $\rm\AA$ with $\mu_0 H$
in T. This empirical relationship is plotted as a dotted line in
Fig.~\ref{fig:field}. One can see that the CO correlation length
does not change at small field, but starts to increase when the
separation between vortices become similar to the CO correlation
length. Therefore, our observation is strongly suggestive of the
role played by vortices in determining the stripe correlation
length. Further theoretical calculation would be required to
elucidate microscopic picture of vortices in stripe ordered
cuprates.

We note that, in their study of the transport properties of $\rm
La_{15/8}Ba_{1/8}CuO_4$ ($\rm T_c$=2.4~K), \cite{QLi07} Li and
coworkers observed that the in-plane resistivity drops rapidly below
about 16~K in zero field, which was interpreted as the onset of
two-dimensional (2D) superconducting transition. Above this
temperature, the unbinding of thermally excited vortex-antivortex
pairs creates phase fluctuations, and the 2D superconductivity is
lost. They also observed that this 2D superconducting transition
temperature is suppressed with the c-axis magnetic field, and
extrapolates to about 4-5~T at zero temperature, which is similar to
the ``critical field" in our study. Since the in-plane resistivity
is an indirect measure of inverse superconducting correlation
area,\cite{Berg07} the data in Ref.~\onlinecite{QLi07} suggests that
the superconducting correlation length actually decreases. It is
interesting to note that this behavior of superconducting
correlation length is opposite of what is observed for stripe
correlation length in our experiment.

An alternative way to describe stripes and superconductivity in LBCO
is the competing order picture. In its extreme form, the competing
charge and superconducting order will result in microscopic phase
separation. Specifically, one can imagine that the charge order and
the superconducting order exist in spatially distinct regions of the
sample, and competition between these phases drive the subtle change
in their respective domain size. This picture of microscopic phase
separation was suggested by Uemura \textit{et al.} based on their
muon spin rotation ($\mu$SR) study of the Zn substituted cuprates.
\cite{Nachumi96,Uemura04} In their recent $\mu$SR investigation of
$\rm (La,Eu,Sr)_{2}CuO_4$, Kojima and coworkers proposed that static
stripe magnetism occurs in different spatial regions than the
superconductivity. \cite{Kojima03}

In summary, we have observed that the correlation length of the
charge order in the superconducting phase of $\rm
La_{2-x}Ba_{x}CuO_4$ ($\rm x\approx1/8$) increases as the magnetic
field greater than $\sim 5$~T is applied. This ''critical" field
scale of $\sim 5$~T seems to be an important point in the phase
diagram. At this point, the vortex lattice spacing becomes
comparable to charge stripe correlation length, and the 2D
superconductivity is lost due to phase fluctuations. Taken together,
our study suggests that static charge order competes with the
superconducting ground state in this LBCO sample. Our experiment
also demonstrates that the field-dependent x-ray scattering
experiment is useful in probing the physics of charge order and
superconductivity.

\acknowledgements{We would like to thank E. A. Kim, S. Kivelson, S.
Sachdev, and J. Tranquada for invaluable discussions, S. LaMarra for
the work at NSLS. Research at the University of Toronto was
supported by the Natural Sciences and Engineering Research Council
of Canada. Work at Brookhaven was supported by the U. S. DOE, Office
of Science Contract No. DE-AC02-98CH10886. Use of the Advanced
Photon Source was supported by the U. S. DOE, Office of Science,
Office of Basic Energy Sciences, under Contract No.
W-31-109-ENG-38.}

%\bibliography{a-d,e-j,k-r,s-z,mypaper}
\bibliography{lbco}

\begin{thebibliography}{33}
\bibitem[{His()}]{History}
\bibinfo{note}{In this study, we have not systematically investigated the
  history dependence.}
\end{thebibliography}

\end{document}